     \documentstyle[12pt,epsfig,wrapfig]{article}   
     \newlength{\dinwidth}                       
     \newlength{\dinmargin}                      
\newcommand{\xb}{\mbox{$x~$}}  % Bjorken x
\newcommand{\Qsq}{\mbox{$Q^2~$}}

\newcommand{\GeV}{\mbox{\rm ~GeV~}}
\newcommand{\GeVx}{\rm GeV}

\newcommand{\GeVsq}{\mbox{${\rm ~GeV}^2~$}}
\newcommand{\GeVsqx}{\mbox{${\rm ~GeV}^2$}}

\newcommand{\ep}{\mbox{$e^{\pm}p~$}}

\newcommand{\als}{$\alpha_s$}

\def\lsim{\mathrel{\rlap{\lower4pt\hbox{\hskip1pt$\sim$}}
    \raise1pt\hbox{$<$}}}                % less than or approx. symbol
\def\gsim{\mathrel{\rlap{\lower4pt\hbox{\hskip1pt$\sim$}}
    \raise1pt\hbox{$>$}}}                % greater than or approx. symbol

\begin{document}
\bibliographystyle{jetnotit}
\vspace*{-18mm}
{\normalsize \rightline{MPI-PhE/99-03}
\rightline{hep-ph/9906541}} 
%
%\input titelblatt
%\vspace*{-8mm}
\begin{center}  \begin{Large} \begin{bf}
Renormalisation Scale Dependencies in Dijet Production at HERA \\
  \end{bf}  \end{Large}
  \vspace*{5mm}
  \begin{large}
T. Carli
  \end{large} 
Max-Planck-Institut f\"ur Physik, M\"unchen, Germany\footnote{
Contribution to 
the Proceedings of the DESY Workshop 1998/99 on Monte Carlo Generators for 
HERA Physics.}
\end{center}
%\vspace{-0.5cm} 
%\begin{center}
%\end{center}
%
\begin{quotation}
\vspace{-0.3cm} 
\noindent
{\bf Abstract:}
 Different choices of the renormalisation scale ($\mu_{ren}$) 
 can be used to describe hard scattering processes with two jets 
 at large transverse momentum in deep-inelastic scattering at HERA
 by fixed order perturbative QCD calculations. 
 For leading (LO) and next-to-leading order (NLO) calculations
 the simplest choices, $Q^2$, the virtuality of the
 incoming photon, and the mean squared transverse momenta of the
 two jets $E_T^2$, are studied in different kinematic regimes.
 It is found that
 only if both $Q^2$ and $E_T^2$ are large, the NLO 
 calculation is stable with respect to numerical variations
 of $\mu_{ren}$ while the LO calculation strongly depends
 on it. If only one of the two scales, either $Q^2$ or $E_T^2$,
 is large, the NLO is more stable than the LO calculation, 
 but exhibits nevertheless a strong residual scale dependence.
 When both scales are relatively small, large scale dependencies
 are found in both cases. Moreover, large differences between
 the LO and NLO calculation are found.

 Generally, the use of $E_T^2$ as renormalisation scale is
 favoured over $Q^2$, since scale dependencies are less
 pronounced and NLO corrections are smaller.

\end{quotation}
\vspace{-0.5cm} 
\section{Introduction}
HERA colliding $27.5$~\GeV positrons on $820$~\GeV proton
offers  an ideal testing ground for perturbative
QCD (pQCD) in deep-inelastic scattering (DIS). 
The centre of mass energy of 
$\sqrt{s} \approx 300 \GeV$ leads to a large phase space for hadron
production and to the possibility to observe
collimated sprays of hadrons - called ´jets´ - 
in the hadronic final state. 
They are the experimentally accessible signs
of a hard scattering process and
relate the fundamental objects of pQCD,
the unobservable quarks and gluons, to the measurable 
hadronic final state. 

\begin{figure}
%\rule{5cm}{0.2mm}\hfill\rule{5cm}{0.2mm}
%\vskip 2.5cm
%\rule{5cm}{0.2mm}\hfill\rule{5cm}{0.2mm}
%\hspace{-1cm}
\epsfig{figure=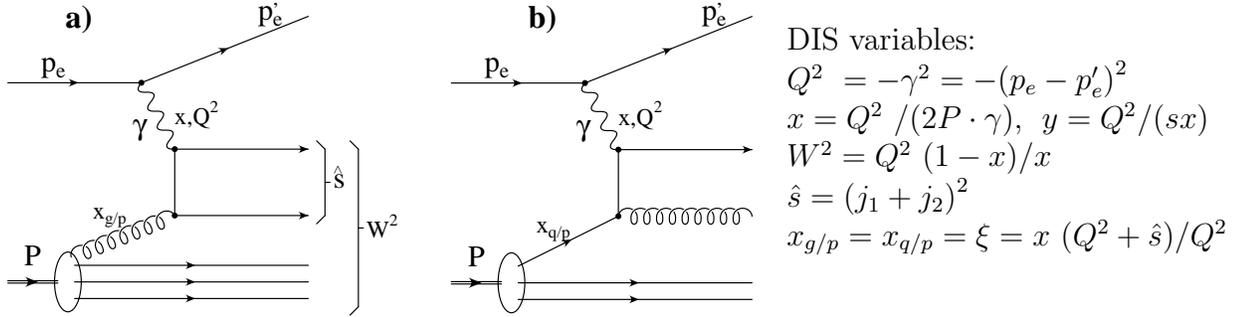,width=10cm}
%        bbllx=401,bblly=729,bburx=137,bbury=323
%        ,width=8cm,angle=270,clip=}
\begin{tabular}{l}
 \vspace{-5.cm} \\
 DIS variables: \\
   $\Qsq= - \gamma^2 = - {(p_e - p_e^\prime)}^2$ \\
   $ x = \Qsq / (2 P \cdot \gamma) $,  $\; y = Q^2/(s x) $\\
   $ W^2 = \Qsq (1 - x )/x $ \\
 $\hat{s} = {(j_1 + j_2)}^2$ \\
 $x_{g/p} = x_{q/p}= \xi= \xb (Q^2 + \hat{s})/Q^2$ \\
% $z_q= (P \cdot j_1)/(P \cdot q)$ \\
%}
\end{tabular}

\caption{\it Feynman diagrams for the production of dijet events 
to first order of \als~in \ep-collisions.
$\gamma$ ($P$) denotes the four-momentum of the photon (proton). 
 $j_1$ and $j_2$ are the four-momenta of the jets associated
to the hard subprocess.
\label{fig:feynjets}}
\vspace{-0.5cm}
\end{figure}

At HERA, events with two jets in the central part of the detector
%plus the jet associated with the proton remnant 
can be produced in quark ($q\gamma \to q g$) 
or gluon ($g\gamma \to q \bar{q}$) 
initiated hard subprocesses (see Fig.~\ref{fig:feynjets}).
In fixed order pQCD the dijet cross section 
can be written as: 
%%%%%%%%%%%%%%%%%%%%%%%%%%%%%%%%%%%%%%%%%%
\begin{eqnarray}
\vspace{-0.2cm}
\frac{d^2\sigma_{dijet}}{dx dQ^2} \sim 
\sum_n \alpha_s^n(\mu^2_{ren}) \;
\int_0^1 \frac{d\xi}{\xi} \;
%\left (
%(
C_n(x/\xi,\mu^2_{ren},\mu^2_{fac},...) 
\cdot PDF(\xi,\mu_{fac}^2)
%+ %\nonumber \\
%C_{q\bar{q}}(x/\xi,Q^2/\mu^2_{ren},\mu^2_{fac}) 
%\cdot g(\xi,\mu_{fac}^2) 
%)
% \right )
%\vspace{-0.2cm}
\label{eq:cross}
\end{eqnarray}
%%%%%%%%%%%%%%%%%%%%%%%%%%%%%%%%%%%%%%%%%%
The variables \xb and \Qsq are the usual variables to 
inclusively describe DIS (see Fig.~\ref{fig:feynjets} 
for definition). 
The (non-perturbative) universal parton density 
functions of the incoming proton $PDF$ are evaluated
at the factorisation scale $\mu^2_{fac}$.
They further depend on $\xi = x \; ( 1 + \hat{s}/Q^2)$, where
$\sqrt{\hat{s}}$ is the invariant mass of the hard partonic system. 
The variable $\xi$ can be interpreted as the
fractional momentum of the parton initiating the
hard subprocess with respect to the proton momentum.
In leading order, where $\hat{s}$ is the invariant mass of the
dijet system, $\xi$ can be directly measured once the dijet
system is tagged.
The coefficient functions $C_n$ 
can be calculated in pQCD as power series expansions
in the strong coupling constant $\alpha_s(\mu^2_{ren})$ probed
at the squared renormalisation scale $\mu_{ren}^2$.  
%
%and depend\footnote{The dependence of
%$C_{qg}$ and $C_{q\bar{q}}$ on the 
%azimuthal angle $\phi$ between the lepton and the parton
%scattering plane and on 
%$z_q = (P \cdot j_1)/(P \cdot q) 
%\approx 1/2 (1-\cos{\theta^*})$ where
%$\theta^*$ is the polar angle
%of one of the jets with $4$-momentum $j_1$
%in the photon-parton center of mass system,
%is not given in the formula.} on the renormalisation 
%and the factorisation scale $\mu_{fac}$ as well as on 
%characterizing the short distance subprocess which is
%responsible for the dijet system.

%%%%%%%%%%%%%%%%%%%%%%%%%%%%%%
%%%%%%%%%%%%%%%%%%%%%%%%%

The dijet cross section including
higher order parton emissions are difficult to calculate using 
exact expressions for the coefficient functions. 
Since the phase space integrals cannot be solved analytically,
numerical methods have to be applied.
Several Monte Carlo integration programs are available
to compute jet cross sections in next-to-leading order (NLO)
of \als
(MEPJET \cite{jet:mirkes96c,jet:mirkes96d}, 
JETVIP \cite{jet:jetvip},
DISENT \cite{jet:sey96d,jet:sey96e} and 
DISASTER \cite{jet:graudenz97a}).
They have been carefully compared in this workshop \cite{jet:wobisch99}. 
They allow any jet definition scheme and
arbitrary experimental cuts to be analysed.
In this article all results are based on the DISENT program.
NLO (LO) calculations are performed using the
CTEQ4M (CTEQ4L) parton density functions \cite{th:cteq4}.
The value of $\alpha_s$ is taken from the parton density functions.
In CTEQ4L $\alpha_s(M_Z)= 0.131$ and 
in CTEQ4M $\alpha_s(M_Z)= 0.116$  is used. 
The factorisation scale has been set to $\mu^2_{fac}= Q^2$.
The dependencies of the results presented here
on the factorisation scales are small and can be neglected. 

\section{Choice of the Renormalisation Scale in DIS}
%%%%%%%%%%%%%%%%%%%%%%%%%%%%%%
%\begin{figure}
\begin{wrapfigure}[16]{l}{8.3cm}
\vspace{-0.5cm}
\epsfig{figure=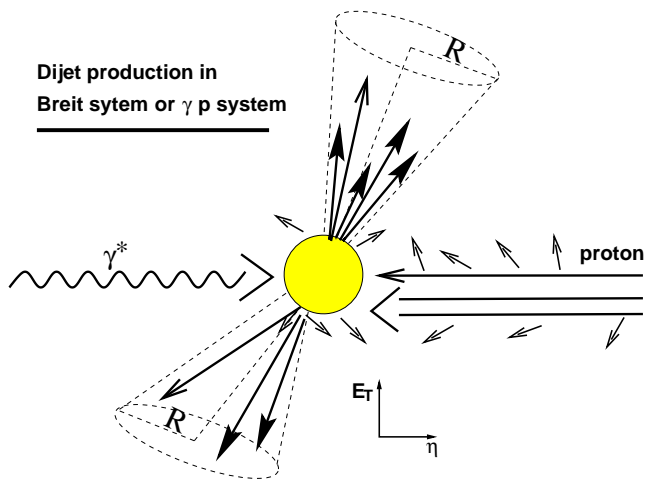,width=8.5cm}
\caption{\it Diagram for the production of dijet events
in the Breit or hcms frame in DIS. 
\label{fig:2jets}}
\end{wrapfigure}
%\vspace{0.2cm}
%\end{figure}
%%%%%%%%%%%%%%%%%%%%%%%%%
%

In pQCD calculations to all orders of $\alpha_s$ the dijet
cross section does not depend on the unphysical
quantities $\mu_{ren}$ and $\mu_{fac}$. However, in a
calculation to finite order a residual scale dependence
is expected. The size of the scale dependence
can be used to estimate the
higher order contributions which have been neglected
in the calculation. It is expected that they are smallest
when the perturbative expansion is performed at the
``natural'' scale. 
 In DIS jet production is a two-scale problem where it is
 not a priori clear at which scale $\alpha_s$  and the
 parton density function should be probed.
 In a frame\footnote{e.g. the hadronic centre of mass
 (hcms) frame defined by $\vec{\gamma} + \vec{P} = 0$ or the
 Breit frame defined by $\vec{\gamma} + 2 \xb \vec{P} = 0$.
 In the laboratory frame and in the Breit frame the proton
 moves towards the $+z$ direction.},
 where the virtual photon and the proton collide
 head-on, it is obvious that two scales can characterise
 the hardness of the process: $Q^2$, the virtuality of the
 incoming photon, as used in the QCD analysis of the
 inclusive DIS cross section, or the mean squared transverse momenta of the
 two jets $E_T^2$ as used in hadron-hadron collisions.
 In this article the behaviour of the dijet cross section
 in different kinematic regimes for both $Q^2$ and $E_T^2$
 as nominal renormalisation scales is investigated.

 To decide whether a physical process is probed at its natural scale
 and whether the pQCD prediction can be expected to be stable,
 the following criteria could give some guidance.
 The NLO corrections, i.e. the ratio of the NLO to the LO
 cross section, the so-called ´K-factor´, should not be
 too large, since in this case it is conceivable that
 corrections of the next-to-next-to-leading order can be
 large. The residual scale dependence, i.e. the variation
 of the cross section when changing the nominal squared renormalisation scale
 ($\mu_{nom}^2$) by about an order of magnitude using 
 a scale factor $1/4 \le \xi_{ren} \le 4$, 
 should be small ($\mu^2_{ren}=\xi_{ren} \cdot \mu^2_{nom}$). 
% Moreover, there should be some improvement in case of the
% NLO calculation when $\xi_{ren}$ is varied.
 Moreover, the NLO calculation should depend less on variations
 of $\xi_{ren}$.

 \vspace{-0.4cm}
 \section{Dijet Rates and Cut Scenario}
%%%%%%%%%%%%%%%%%%%%%%%%%%%%%%%%%%%%%%%%%%%%%%%%%%%%%%%%%%%
\begin{figure}[htb] 
  \vspace*{-18mm}
%     kumac: rec06:~/jet/delta#nice
 \begin{center}
 \epsfig{figure=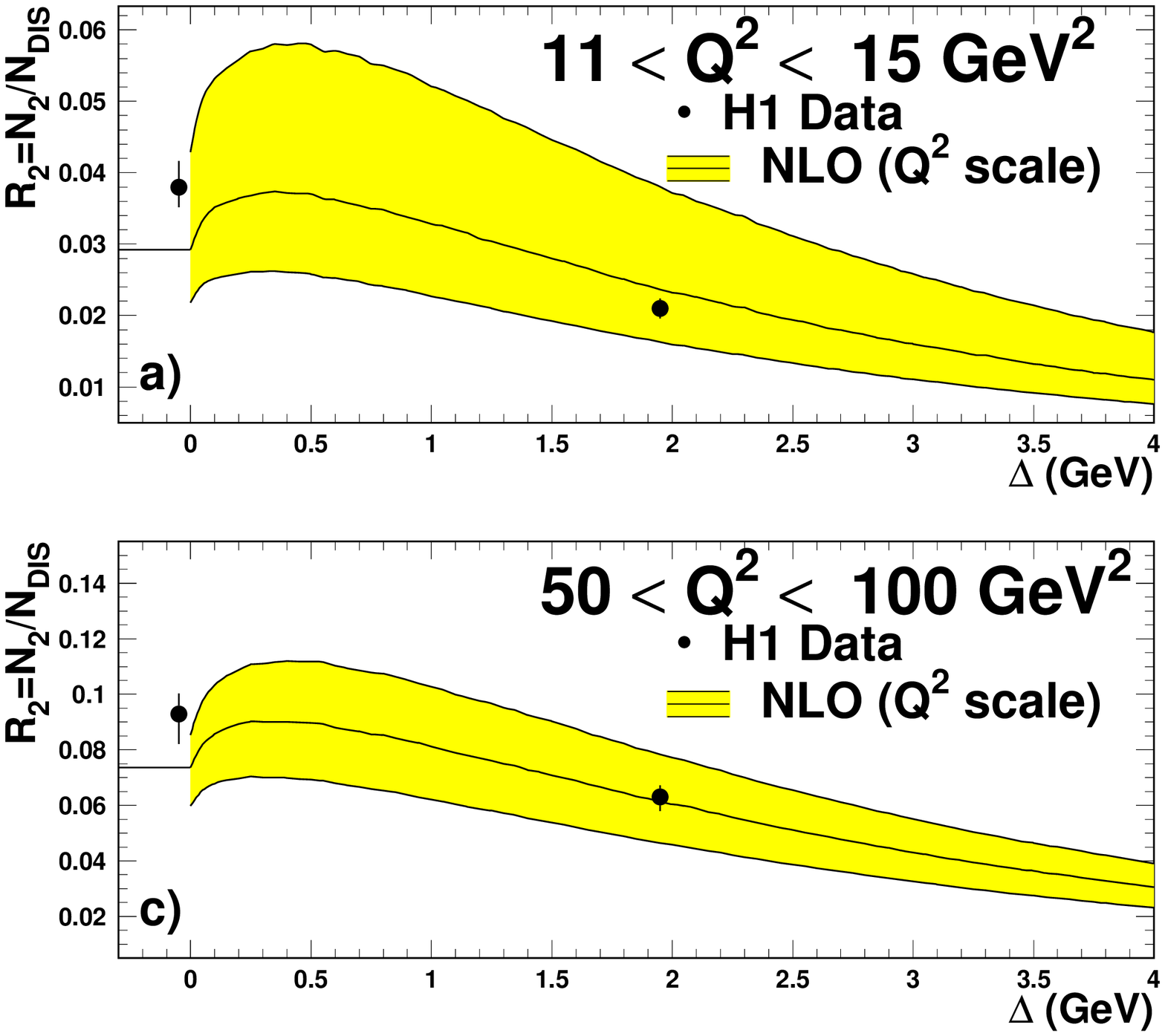,width=0.48\textwidth}
 \epsfig{figure=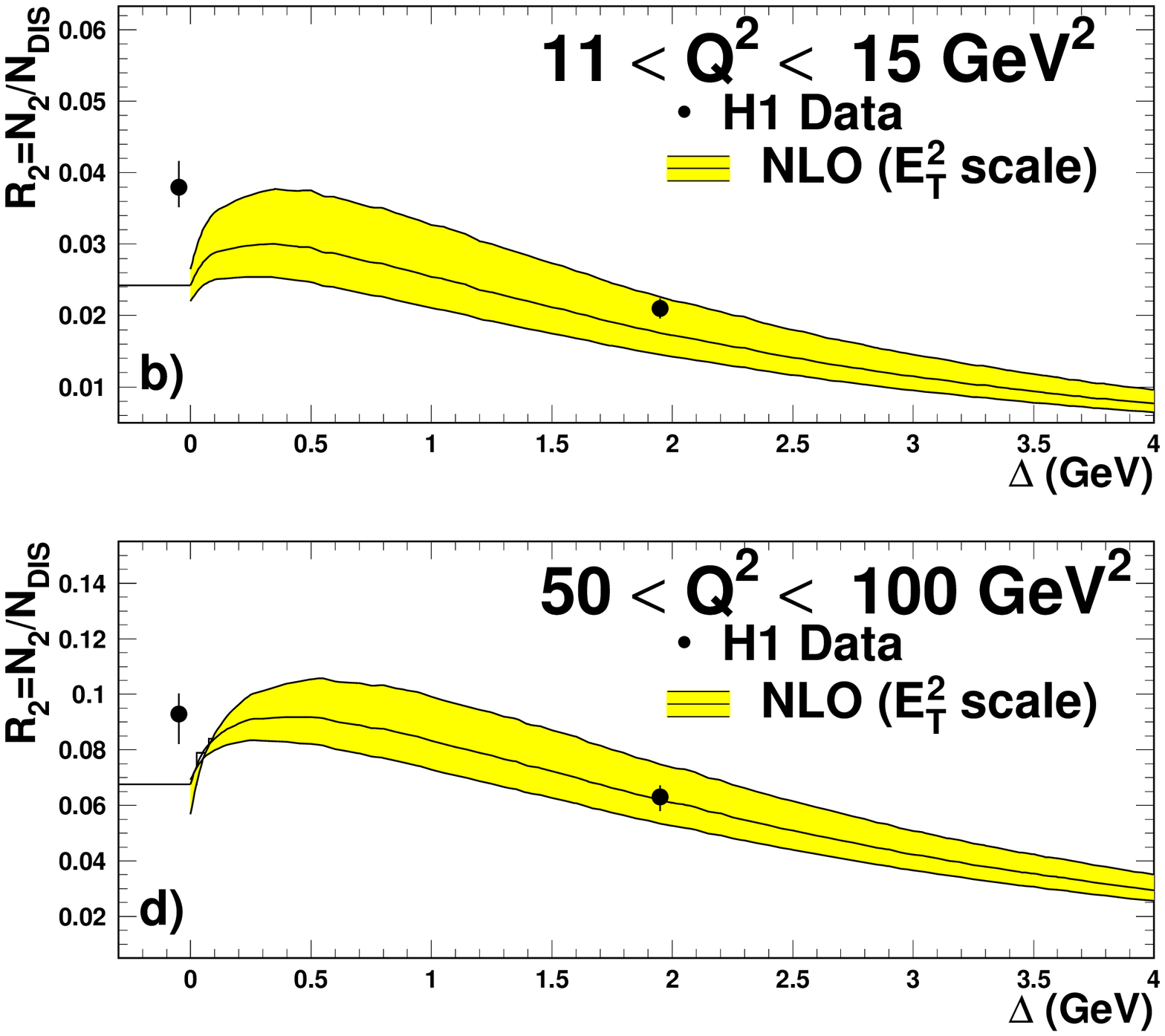,width=0.48\textwidth}
 \end{center} 
\vspace{-0.8cm}
\caption[$R_2$ as a function of $\Delta$]{\label{fig:delta} {\it
 Dijet rate $R_2 = N_{2}/N_{DIS}$ as a function of the
 threshold difference $\Delta$ of the first and second jet.
 Jets are defined by the cone algorithm ($R_{cone}=1$)
 in the hadronic centre of mass system with 
 $E_{T2} \ge 5$ \GeV and  $E_{T1} \ge 5 + \Delta$ \GeVx.
 {\it Shown are H1 data (points) for $11 \le Q^2 \le 15$ \GeVsq
 (a,b) and $50 \le Q^2 \le 100$ \GeVsq (c,d) together with
 a NLO QCD calculation (line) with 
 $Q^2$ (a,c) and the mean $E_T^2$ (b,d) as renormalisation 
 scale $\mu^2_{nom}$. 
 The shaded band indicates the variation of the NLO prediction
 for $4 \cdot \mu^2_{nom}$ and  $1/4 \cdot \mu^2_{nom}.$} 
  }}
\end{figure}
%%%%%%%%%%%%%%%%%%%%%%%%%%%%%%%%%%%%%%%%%%%%%%%%%%%%%%

 Before studying the scale dependencies of the dijet
 cross section,
 it is first indispensable to find a cut scenario
 where pQCD can describe the data. For dijet production in DIS
 this has been a problem at HERA. 
 For a long time agreement with NLO
 calculations could only be achieved in very
 restricted phase space regions (see e.g.~\cite{jet:ringberg}).
 Especially in the region of low $Q^2 \lsim 50$ \GeVsqx, 
 NLO calculations 
 %and LO QCD models, e.g. like implemented in 
 %LEPTO\cite{mc:lepto} or HERWIG\cite{mc:herwig},
 failed to describe the data. 
 Meanwhile, it has been understood that this
 discrepancy occurs when both jets are required
 to be above the same $E_T$ threshold.
 As $\Delta$ defined as the required $E_T$ difference of the threshold
 of the two leading jets, approaches $0$, a fixed order calculation gets 
 infra-red sensitive~\cite{jet:frix97,jet:kramer96,jet:potter98}.
 In this region a reliable prediction of the jet cross
 section is not possible with a fixed order calculation.

 This has been demonstrated in an analysis by the H1
 collaboration~\cite{jet:h1dijet},
 where $R_2 = N_2/N_{DIS}$, the ratio 
 of the two jet cross section to the inclusive DIS 
 cross section, has been measured in the region\footnote{
 $\theta_{el}$ ($E_{el}$) is the polar angle (energy) of the 
 scattered electron.} $5 \le Q^2 \le 100$ \GeVsqx,
 $156^o \le \theta_{el} \le 173^o$, $E_{el} \ge 11$ \GeV
 and $y \ge 0.05$. Jets are defined in the hadronic
 centre of mass system by the cone 
 algorithm~\cite{jet:cdf_92} with a radius $R_{cone}=1$. 
 The $E_T$ of the second jet is required to be
 $E_{T2} \ge 5$~\GeV and the $E_T$ of the first jet
 $E_{T1} \ge 5 + \Delta$~\GeVx, where $\Delta$ is set to
 $0$ or $2$ \GeVx.
 The difference of their pseudo-rapidities\footnote{
 The pseudo-rapidity is defined as 
 $\eta = - \ln{\tan{\theta/2}}$. In the laboratory and in the
 Breit frame the proton moves into the $+z$ direction.} 
 should be $|\eta^*| < 2$.

%%%%%%%%%%%%%%%%%%%%%%%%%%%%%%%%%%%%%%%%%%%%%%%%%%%%%%%%%%%
\begin{figure}[htb] 
  \vspace*{-10mm}
%     kumac: rec06:~/jet/r2 
 \begin{center}
 \epsfig{figure=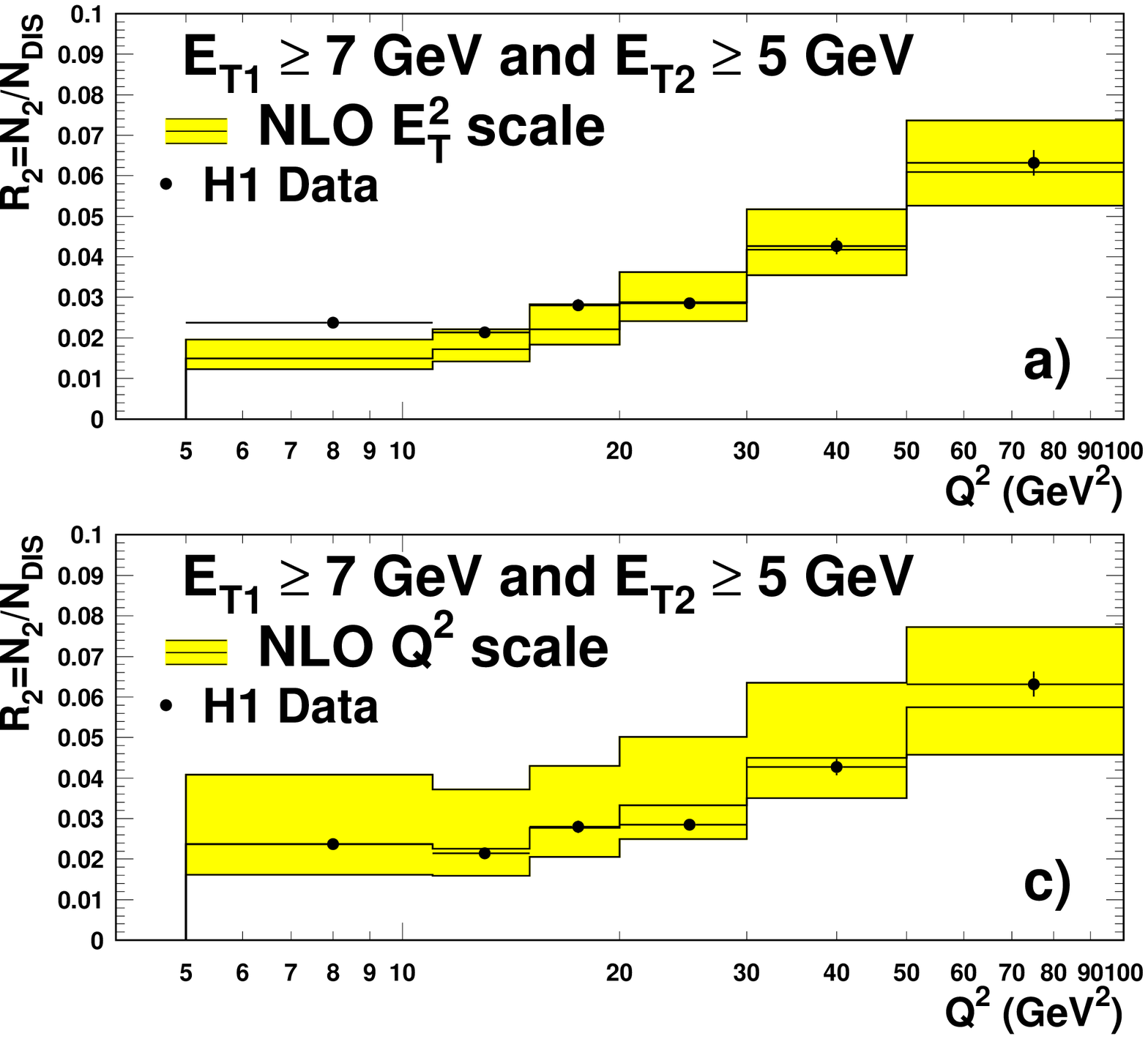,width=0.48\textwidth}
 \epsfig{figure=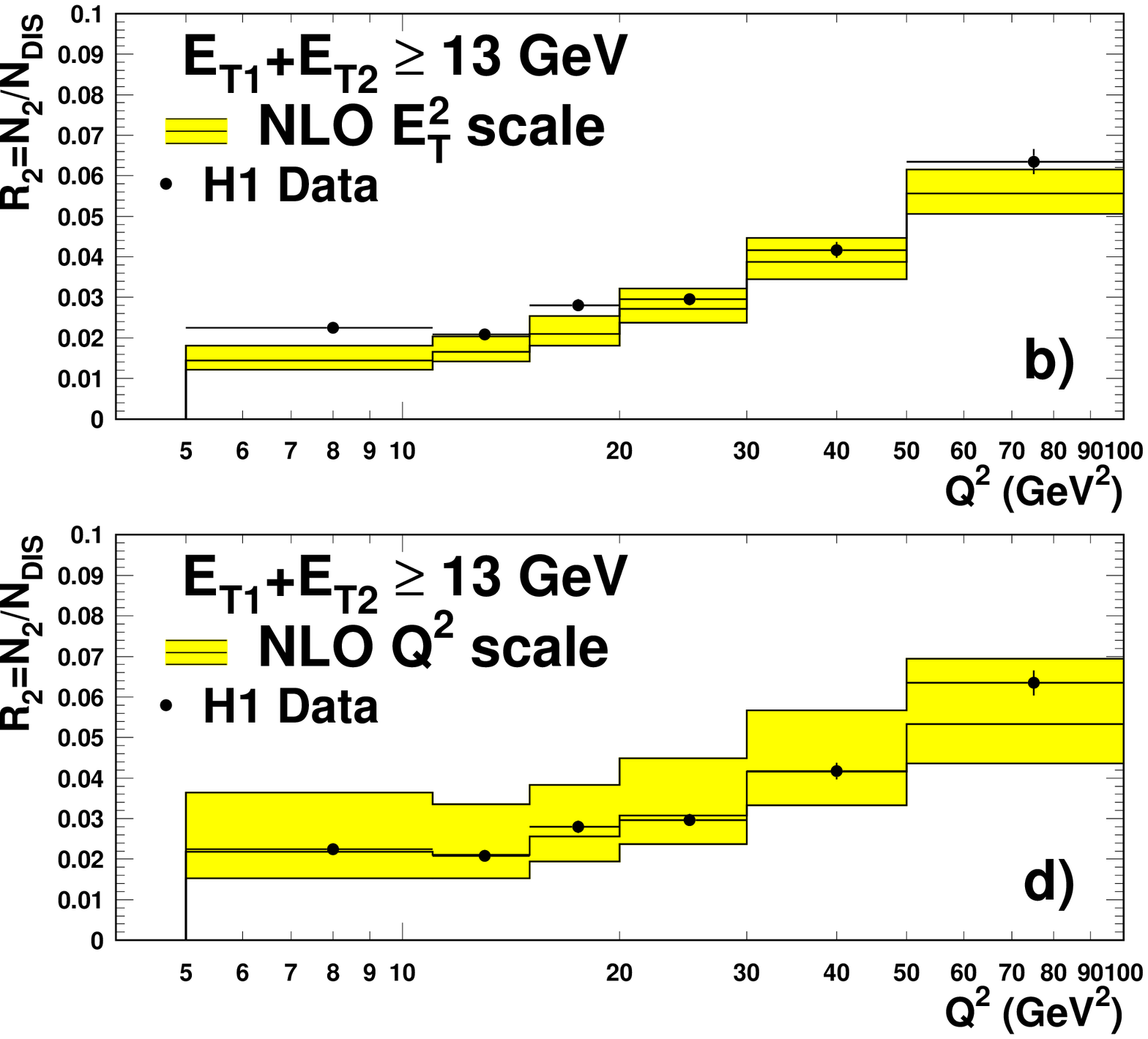,width=0.48\textwidth}
 \end{center} 
\vspace{-0.8cm}
\caption[$R_2$ as a function of $Q^2$]{\label{fig:r2vsq2}{\it
 Dijet rate $R_2 = N_{2}/N_{DIS}$ as a function of $Q^2$.
 Jets are defined by the cone algorithm ($R_{cone}=1$)
 in the hadronic centre of mass system with 
 $E_{T2} \ge 5$ \GeV and  $E_{T1} \ge 7$ \GeV (a,c)
 or
 $E_{T2} + E_{T1} \ge 13$ \GeV (b,d).
 Shown are H1 data (points) together with
 a NLO QCD calculation (line) with 
 $Q^2$ (c,d) and the mean $E_T^2$ (a,b) as renormalisation 
 scale $\mu^2_{nom}$. 
 The shaded band indicates the variation of the NLO prediction
 for $4 \cdot \mu^2_{nom}$ and  $1/4 \cdot \mu^2_{nom}$.  
  }}
\end{figure}
%%%%%%%%%%%%%%%%%%%%%%%%%%%%%%%%%%%%%%%%%%%%%%%%%%%%%%

 The data points for $11 \le Q^2 \le 15$ \GeVsq (a,b)
 and $50 \le Q^2 \le 100$ \GeVsq (c,d) are shown in
 Fig.~\ref{fig:delta}. As expected, in the data the dijet rate is
 higher at $\Delta=0$ \GeV than it is at
 $\Delta=2$ \GeVx. In the NLO calculation the expected
 rise of the dijet rate towards decreasing $\Delta$
 is also visible. However, around 
 $\Delta \approx 0.5$ \GeV the calculated dijet rate
 turns around and falls down until $\Delta \approx 0$ \GeV 
 is reached. Around this point the NLO cross section falls
 with an infinite slope \cite{jet:frix97}, but remains
 finite.
 In this region the $E_T$ of the jets
 are approximately equal and no phase space is available
 to emit a third real parton. This leads to an 
 incomplete cancellation between real and virtual
 corrections at the threshold and makes a fixed
 order calculation unpredictive. 
 A resummation of higher
 orders is necessary at this phase space point.
 Since the virtual corrections give a negative
 contribution, the dijet rate artificially drops
 down in the NLO calculation.
 The problem only occurs for events containing two jets
 with balanced momenta. In such a configuration the
 emission of a real third parton would lead to a configuration
 where one parton is above the $E_T$ threshold and
 the other below. If the thresholds of the two
 jets are different, the event has either balanced
 jet momenta or has jets with one above and the other
 below the thresholds. Since these conditions are never
 fulfilled at the same time, no problem occurs.

 In Fig.~\ref{fig:delta} one can see that the point
 $\Delta = 2$ \GeV is well described by the calculation
 while for $\Delta = 0$ \GeV the calculation is below
 the data.
 This behaviour is independent of the used renormalisation scale.
 However, when $Q^2$ is used a renormalisation scale
 a much larger scale dependence is found.
 In Fig.~\ref{fig:r2vsq2}a and \ref{fig:r2vsq2}c 
 it is demonstrated that
 the NLO calculation agrees well with the data for 
 $5 \le Q^2 \le 100$ \GeVsqx, if $\Delta = 0$ \GeV is avoided.
 Hadronisation corrections would lower the
 NLO prediction by about $10-20\%$.
 Only in the case of the lowest $Q^2$ bin the NLO calculation
 falls below the data. 
 The difficult phase space region can not only be avoided by a cut
 on $\Delta$, but also by a cut on the sum of the two jets, e.g.
 $E_{T1} + E_{T2} \ge 13$ \GeV     
 (see Fig.~\ref{fig:r2vsq2}b and \ref{fig:r2vsq2}d).   
 For both cut scenarios the $Q^2$ scale exhibits a much larger
 scale dependence than the $E_T^2$ scale 
 (where $E_T = (E_{T1}+E_{T2})/2$).
 The variation of the NLO calculation when changing the
 renormalisation scale by a factor of $4$ up and down
 is illustrated as grey band.
 The central value for a scale factor of $\xi_{ren} = 1$ is 
 given as solid line.
 For both choices of scales, $Q^2$ and $E_T^2$, the cut on the 
 sum of the two jets
 leads to a slightly smaller scale dependence than the
 $\Delta=2$ \GeV cut.

 \vspace{-0.5cm}
 \section{Dijet Cross Sections and their NLO Behaviour}
%%%%%%%%%%%%%%%%%%%%%%%%%%%%%%%%%%%%%%%%%%%%%%%%%%%%%%%%%%%
\begin{wrapfigure}[33]{l}{9.5cm}
%\begin{figure}[htb] 
  \vspace*{-15mm}
%     kumac: rec06:~/jet/cone
% \begin{center}
 \epsfig{figure=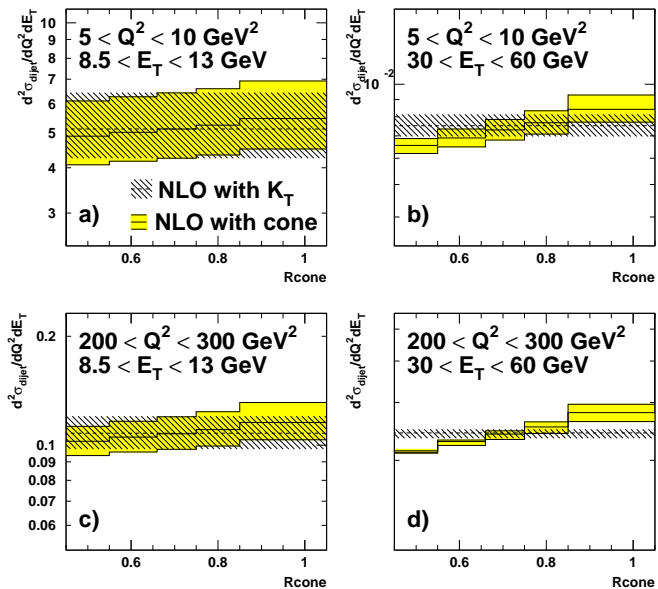,width=9.cm}
% \end{center} 
\vspace{-0.8cm}
\caption[$R_{cone}$ dependence of dijet cross section]{\label{fig:conevskt}
{\it
Dependence of the double differential dijet cross section 
$d^2\sigma_{dijet}/dQ^2 dE_T$ defined by the
cone algorithm with a cone size of $R_{cone}$ as calculated
by NLO QCD ($E_T^2$ scale) in different
bins of $Q^2$ and of the mean $E_T$ in the Breit frame.
The shaded band indicates the variation of the NLO prediction
for $4 \cdot \mu^2_{nom}$ and  $1/4 \cdot \mu^2_{nom}$.   
Overlayed as hatched band is the cross section defined
by the inclusive $K_T$ algorithm always for $R=1$. 
Jets are defined
in the Breit frame with $E_{T1} \ge E_{T2} \ge 5$ \GeVx,
$E_{T1} + E_{T2} \ge 17$ \GeV
{\it and $-1 \le \eta_{lab} \le 2.5$.
  }}}
\end{wrapfigure}
%\end{figure}
%%%%%%%%%%%%%%%%%%%%%%%%%%%%%%%%%%%%%%%%%%%%%%%%%%%%%%
 To investigate the interplay between the $Q^2$ and the $E_T^2$
 scales in more detail, the following cut scenario
 is adopted.
 All jets lying well within the detector acceptance
 $-1.5 \le \eta_{lab} \le 2.5$ are ordered in energy.
 Then the two highest $E_T$ jets are required to fulfil:
 $E_{T1} \ge E_{T2} \ge 5$ \GeV and
 $E_{T1} + E_{T2} \ge 17$ \GeV. Jets are defined by the
 inclusive $K_T$ algorithm~\cite{jet:inclkt,jet:cadosewe93} 
 in the Breit system.  
 The kinematic phase space
 is defined by $0.2 \le y \le 0.6$ and $5 \le Q^2 \le 600 $ \GeVsqx.
 These cuts are very close to the ones used by the
 H1 collaboration in a recent extraction of the
 gluon density~\cite{jet:h1dijetgluon98}.

%%%%%%%%%%%%%%%%%%%%%%%%%%%%%%%%%%%%%%%%%%%%%%%%%%%%%%%%%%%
\begin{figure}[htb] 
  \vspace*{-10mm}
%     kumac: rec06:~/jet/kfactor
 \begin{center}
 \epsfig{figure=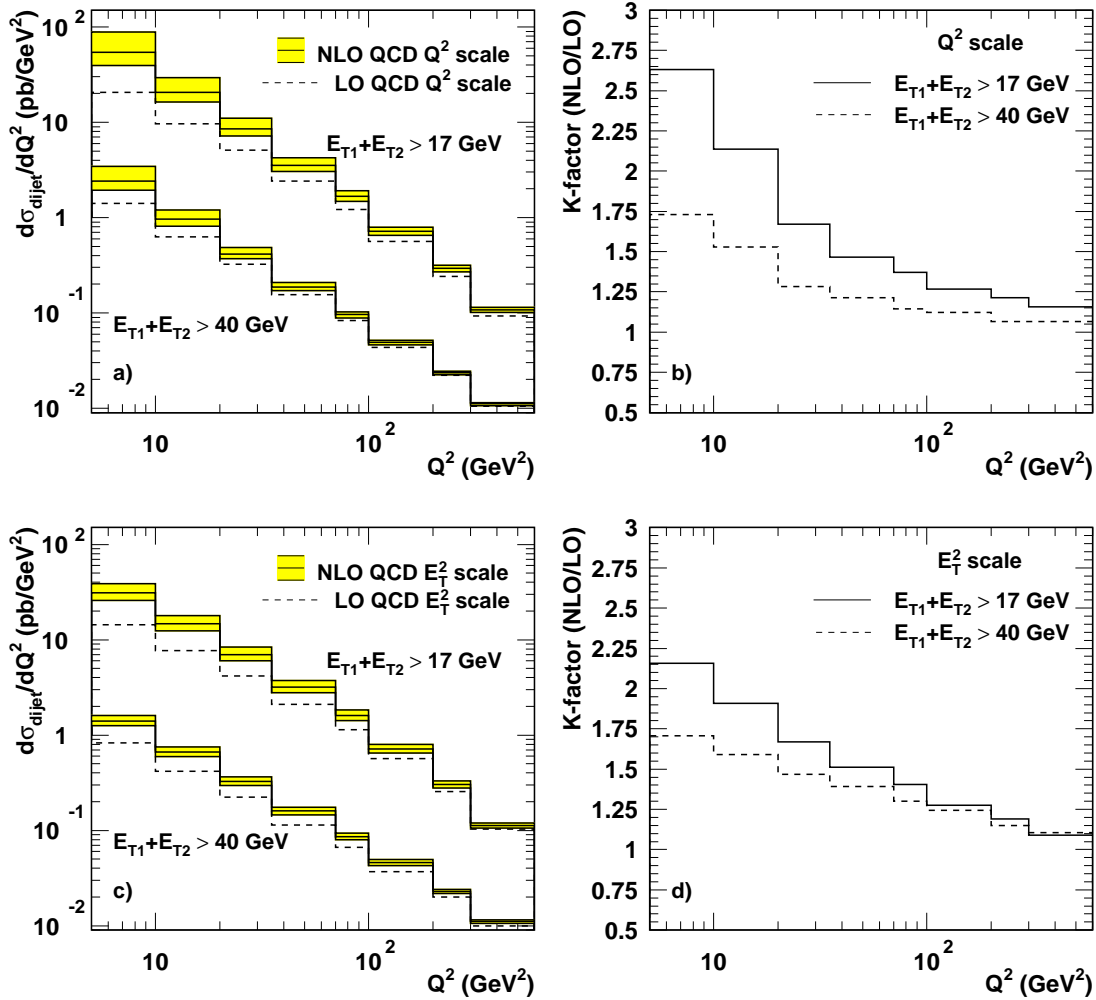,width=0.9\textwidth}
 \end{center} 
\vspace{-1.0cm}
\caption[Dijet cross section as a function of $Q^2$]{\label{fig:dsdq2det2}
{\it
Dijet cross section as a function of $Q^2$ as 
calculated by NLO (line) and LO (dashed line) QCD 
using $Q^2$ (a) 
or the mean $E_T^2$ (c) as renormalisation
scale $\mu_{ren}^2$. 
The shaded band indicates the variation of the NLO prediction
for $4 \cdot \mu^2_{nom}$ and  $1/4 \cdot \mu^2_{nom}$.  
The ratio of the NLO to the LO calculation
is shown in (b) and (d).
Jets are defined by the inclusive $K_T$ algorithm
in the Breit frame with $E_{T1} \ge E_{T2} \ge 5$ \GeV
and  $-1 \le \eta_{lab} \le 2.5$.
Shown are cross sections for 
$E_{T1}+E_{T2} \ge 17$ \GeV
and
$E_{T1}+E_{T2} \ge 40$ \GeVx.
  }}
\end{figure}

%%%%%%%%%%%%%%%%%%%%%%%%%%%%%%%%%%%%%%%%%%%%%%%%%%%%%%

 In a NLO calculation similar results are obtained
 with the inclusive $K_T$ algorithm and with the
 cone algorithm with a cone size $R_{cone}=0.7$.
 This is illustrated in Fig.~\ref{fig:conevskt} where
 the total dijet cross section 
 $d^2\sigma_{dijet}/dQ^2 dE_T$ in two $Q^2$ and mean 
 $E_T=(E_{T,1} + E_{T,2})/2$ 
 bins\footnote{Here and in the following figures, 
 $d^2\sigma_{dijet}/(dQ^2 dE_T)$ is the
 total dijet cross section integrated over a $Q^2$ and $E_T$ 
 bin and divided by the bin width.} 
 is shown as a function of the cone size $R_{cone}$
 used in the cone algorithm\footnote{For the 
 $K_T$ algorithm the distance parameter $R$ is always
 set to $1$. For an exact definition see ref. 
 \cite{jet:h1jetstr}.}. In case of the cone algorithm the scale
 dependence is only slightly influenced by the 
 cone size. Only at large $E_T$ small cone sizes lead
 to improved scale dependencies.
 For $R_{cone}=0.7$ both jet algorithms exhibit the
 same scale dependence. However, the inclusive $K_T$ algorithm
% leads to smaller hadronisation corrections % \cite{XXX???}and 
 is defined in a less ambiguous way and is better suited
 for NLO comparisons with data
 \cite{jet:sey97, jet:kilgore97, jet:Ellis92}.
 Therefore the inclusive $K_T$ jet algorithm is the preferred 
 choice and is used in the following.

 In Fig. \ref{fig:dsdq2det2}a ($Q^2$ scale) and 
 \ref{fig:dsdq2det2}c ($E_T^2$ scale)
 the inclusive dijet cross section
 is shown as a function of $Q^2$
 for $E_{T1} + E_{T2} \ge 17$ \GeV
 and $E_{T1} + E_{T2} \ge 40$ \GeVx.
 At low $Q^2$ where $E_T^2$ is higher than $Q^2$ the corresponding
 cross section calculated with $E_T^2$ as scale
 is lower, since $\alpha_s$ is probed at a higher
 scale. Starting at $Q^2 \gsim 40$ \GeVsq the dijet cross section
 is of comparable size for both scales.
 For both $E_T$ cuts, the scale dependence is largest at lowest
 $Q^2$. The $K$-factor shows a similar behaviour
 (see Fig. \ref{fig:dsdq2det2}b and Fig. \ref{fig:dsdq2det2}d).
 When using $Q^2$ as scale,
 at $Q^2 \approx 10$ \GeVsq for
 $E_{T1} + E_{T2} \ge 17$ \GeV ($E_{T1} + E_{T2} \ge 40$ \GeVx)
 the LO cross section is about
 a factor of $2$ ($1.75$) below the NLO cross section. 
 Only at the largest $Q^2$ of $600$~\GeVsq the NLO correction 
 is small, i.e. about $15\%$ ($5\%$).
 When using $E_T^2$ as scale
 (see Fig. \ref{fig:dsdq2det2}c and Fig. \ref{fig:dsdq2det2}d),
 the scale dependence is
 generally reduced and the K-factor is smaller.

%%%%%%%%%%%%%%%%%%%%%%%%%%%%%%%%%%%%%%%%%%%%%%%%%%%%%%%%%%%
\begin{figure}[htb] 
  \vspace*{-10mm}
 \begin{center}
% kumac: rec06:~/jet/figscale#q2vset
  \epsfig{figure=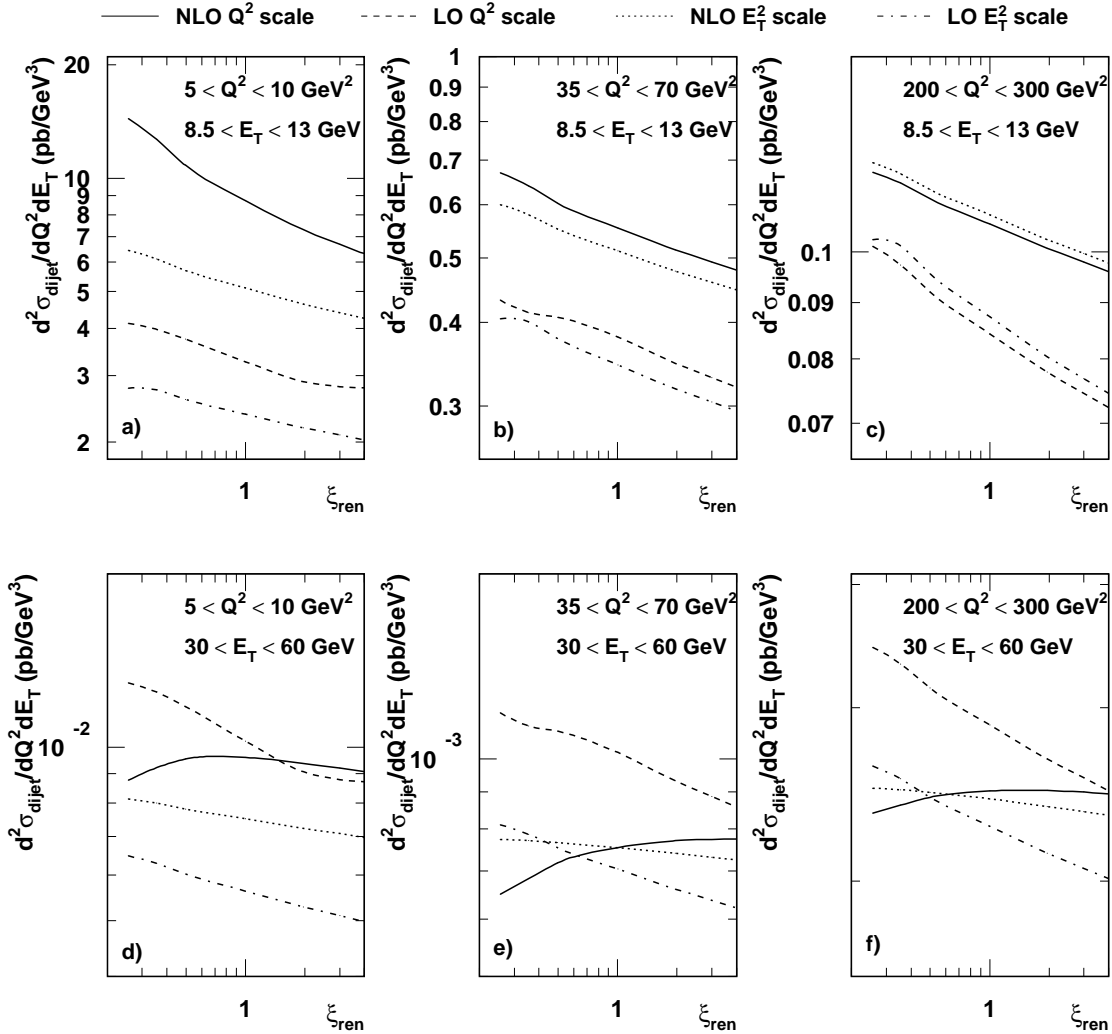,width=0.9\textwidth}  
 \end{center}
\vspace{-1.3cm}
\caption[Double differential dijet cross section 
$d^2\sigma_{dijet}/(dQ^2 dE_T)$ 
as a function of the scale factor $\xi{ren}$]{\label{fig:q2vset}{\it
Double differential dijet cross section 
$d^2\sigma_{dijet}/(dQ^2 dE_T)$ as a function 
of $\xi_{ren}$ 
for different bins of $Q^2$ and the mean $E_T^2$. 
Shown are NLO and LO QCD predictions for $Q^2$ and the mean $E_T^2$ 
as renormalisation scale 
$\mu_{ren}^2 = \xi_{ren} \cdot \mu_{nom}^2$. 
Jets are defined by the inclusive $K_T$ algorithm
in the Breit frame with $E_{T1} \ge E_{T2} \ge 5$ \GeVx,
$E_{T1} + E_{T2} \ge 17$ \GeV
and $-1 \le \eta_{lab} \le 2.5$.
  }}
\end{figure}
%%%%%%%%%%%%%%%%%%%%%%%%%%%%%%%%%%%%%%%%%%%%%%%%%%%%%%

 The explicit dependence of the double differential
 dijet cross section on the scale factor $\xi_{ren}$ is
 illustrated in Fig. \ref{fig:q2vset} for three different
 bins in $Q^2$ and mean $E_T$. Shown are the NLO and LO cross
 sections for $Q^2$ and $E_T^2$ as renormalisation scale.

 For low $Q^2$ ($5 \le Q^2 \le 10$ \GeVsqx) and
 low $E_T$ ($8.5 \le E_T \le 13$ \GeVx) the NLO cross section
 exhibits a strong dependence on $\xi_{ren}$ as is
 shown in Fig. \ref{fig:q2vset}a.
 It varies by a factor $2.5$ for $1/4 \le \xi_{ren} \le 4$ when
 $Q^2$ is used a renormalisation scale. For the $E_T$ scale 
 this dependence is weaker (factor $1.6$ for $1/4 \le \xi_{ren} \le 4$),
 but still large. It is interesting that the LO cross section
 is more stable. It falls only by a factor of $1.5$ for both
 the $Q^2$ and the $E_T^2$ scale.
 When the $Q^2$ scale is replaced by the $E_T^2$ scale,
 the cross section diminishes by a factor of $1.6$ ($1.2$) 
 in case of NLO (LO).
 The K-factors are large, i.e. about a factor of $2 - 2.5$,
 for both scales. The $E_T^2$ scale leads to a slightly smaller
 K-factor.

 Keeping $E_T$ fixed at $8.5 \le E_T \le 13$ \GeV
 and increasing $Q^2$ to $200 \le Q^2 \le 300$ \GeVsqx,
 gives the configuration displayed in 
 Fig. \ref{fig:q2vset}c. Both choices of the renormalisation
 scale result in similar NLO and LO cross sections.
 The cross section is now slightly bigger in case of the
 $E_T^2$ scale. The NLO cross section still reveals a
 distinct dependence on $\xi_{ren}$, but this dependence
 is now weaker than in the case of the LO cross section.
 The K-factors are reduced compared to the low $Q^2$ result,
 but are still large (factor $1.5$). 
 At no value of $\xi_{ren}$ the NLO and the LO cross section
 are similar.

 Fig. \ref{fig:q2vset}d presents the results for
 $5 \le Q^2 \le 10$ \GeVsq and $30 \le E_T \le 60$ \GeVx.
 The dependence of the NLO cross section on $\xi_{ren}$ is
 much reduced for both choices of scales. The LO cross section
 exhibits a much stronger dependence. For large $E_T$
 and $5 \le Q^2 \le 300$ \GeVsq 
 the LO cross section is higher than the NLO cross section
 when $Q^2$ is used as scale. 
 In case of the $Q^2$ scale
 around $\xi_{ren} \approx 2$ the K-factor is $1$. This point
 moves to higher $\xi_{ren}$ values as $Q^2$ increases.
 In case of the $E_T^2$ scale such a point is at very low
 $\xi_{ren}$ values at low $Q^2$ and is around
 $\xi_{ren} = 1/2$ for $35 \le Q^2 \le 300$ \GeVsqx.

 If both $Q^2$ and $E_T$ are large, the NLO cross sections
 are stable for both scale choices (see Fig. \ref{fig:q2vset}f).
 Moreover, the magnitude of the cross section is very
 similar for both scales. The LO cross section is steeply
 falling as $\xi_{ren}$ increases. The K-factor is $1$ for
 $\xi_{ren} = 1/2$ ($\xi_{ren} = 4$) in case of the $E_T^2$ ($Q^2$)
 scale.

\begin{figure}[htb] 
  \vspace*{-10mm}
%     kumac: rec06:~/jet/kfacdetafwd
 \begin{center}
 \epsfig{figure=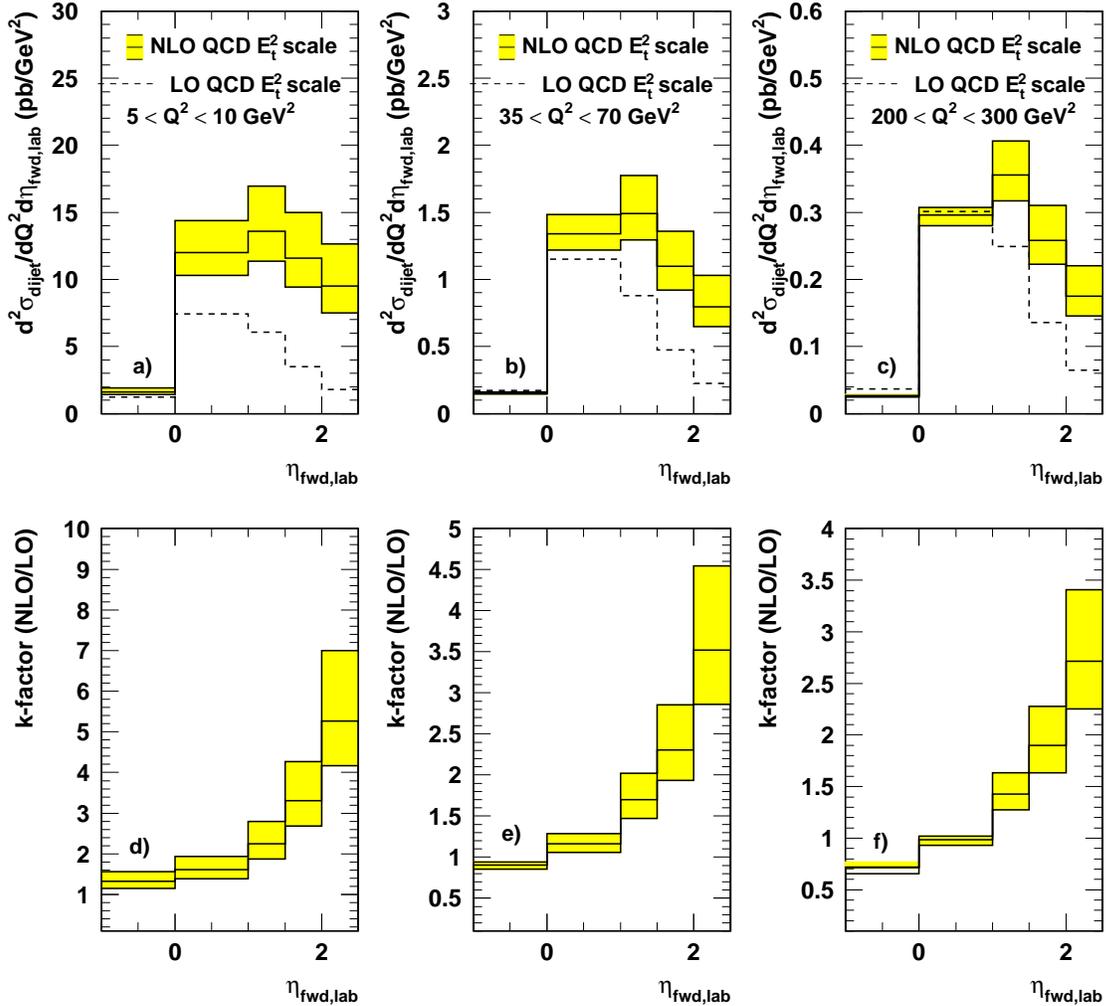,width=0.9\textwidth}
 \end{center} 
\vspace{-1.3cm}
\caption[Double differential dijet cross section 
$d^2\sigma_{dijet}/(dQ^2 d\eta_{fwd,lab}$ 
as a function of $\eta_{fwd,lab}$ ]{\label{fig:kfaceta}
{\it
Dijet cross section 
as a function of the pseudo-rapidity of the forward
jet $\eta_{fwd,lab}$ for different $Q^2$ bins.
Shown are NLO and LO QCD predictions  
for the mean $E_T^2$ as renormalisation scale $\mu_{nom}^2$.
The shaded band indicates the variation of the NLO prediction
for $4 \cdot \mu^2_{nom}$ and  $1/4 \cdot \mu^2_{nom}$.  
Jets are defined by the inclusive $K_T$ algorithm
in the Breit frame with $E_{T1} \ge E_{T2} \ge 5$ \GeVx,
$E_{T1} + E_{T2} \ge 17$ \GeV and
{\it $-1 \le \eta_{lab} \le 2.5$.}
  }}
\vspace{-0.3cm}
\end{figure}

 It is interesting to note that the variation of the NLO
 cross section with the scale factor $\xi_{ren}$ as well
 as the K-factor strongly depend on the position of the jets
 with respect to the proton. This is illustrated in
 Fig. \ref{fig:kfaceta} where the dijet cross section
 as a function of pseudo-rapidity of the forward jet in the
 laboratory frame $\eta_{fwd,lab}$ is shown for three $Q^2$ bins.
 Here $E_T^2$ is used as renormalisation scale. 
 In configurations where both jets are backward, i.e. small
 values of $\eta_{fwd,lab}$, 
 small scale dependencies
 and K-factors around unity are found. However, as the forward
 jet moves towards the proton direction increasingly large scale 
 factors are found.
 For $5 \le Q^2 \le 10$ \GeVsq the NLO correction is $600 \%$ and
 even for  $200 \le Q^2 \le 300$ \GeVsq 
 the NLO correction is still $300 \%$ !
 Also the scale dependence of the NLO cross section
 gradually increases from about $30\%$ for small $\eta_{fwd,lab}$
 to about $70\%$ for large $\eta_{fwd,lab}$ at low $Q^2$.
 Although somewhat smaller, the same effect is visible
 at large $Q^2$.
 These large K-factors suggest that NNLO corrections will be 
 important. In the specific context of forward jet production
 studied in a more restricted phase space, this has already
 been demonstrated~\cite{jet:potterfwdjet}.
 
%%%%%%%%%%%%%%%%%%%%%%%%%%%%%%%%%%%%%%%%%%%%%%%%%%%%%%%%%%%
%%%%%%%%%%%%%%%%%%%%%%%%%%%%%%%%%%%%%%%%%%%%%%%%%%%%%%

% This behaviour of the fixed order calculation indicates
% that higher order parton emissions are expected to be
% more pronounced towards the proton direction.
% This is the region where the influence of initial state
% parton emissions are expected to be most pronounced.
% However, the modelling of these higher orders to a 
% leading logarithm approximation via
% parton showers %\cite{partonshowers} 
% as implemented in the QCD models
% HERWIG \cite{mc:herwig} or LEPTO \cite{mc:lepto} is at low
% $Q^2$ not enough to catch up with the large corrections
% introduced by the ${\cal O}{\alpha_s^2}$ 
% matrix element. For large $Q^2$ they give a fair
% description of these corrections. This is demonstrated
% in Fig. \ref{fig:XXX???}.

 %Other scales: no big difference ET2/KT2

 \vspace{-0.4cm}
 \section{Conclusions}
 \vspace{-0.2cm}

 The behaviour of fixed order calculations
 in dijet production at HERA has been investigated. 
 A cut scenario where the transverse energy of both jets 
 are required to be above the same threshold has to be avoided
 to get agreement of the NLO calculation with the data.
 Dijet rates for $5 \le Q^2 \le 100$ \GeVsq can be described
 by a NLO calculation when using the $E_T^2$ or the $Q^2$
 scale. The $E_T^2$ scale leads to smaller scale dependencies. 

 The interplay between the virtuality of the photon $Q^2$ 
 and the mean transverse energy of the dijet system in the Breit 
 frame $E_T^2$ has been studied in detail using 
 dijet cross sections for $5 \le Q^2 \le 600$ \GeVsq
 and $8.5 \le E_T \le 60$ \GeV as observable.
 
 Only if both $Q^2$ and $E_T^2$ are large, the pQCD seems
 to make reliable predictions. The NLO calculation is
 independent of the renormalisation scale factor 
 $\xi_{ren}$ while the LO calculation strongly depends
 on it. Moreover, a point can be found where the NLO
 and the LO cross section have approximately the same
 size. If either $Q^2$ or $E_T^2$ is large
 and the other scale small, the NLO
 is more stable than the LO calculation, but exhibits
 nevertheless a dependence on $\xi_{ren}$.
 This is more pronounced, if $Q^2$ is large and $E_T^2$ small
 than in the opposite case where 
 $E_T^2$ is large and $Q^2$ small.
 When both scales are relatively small, large scale dependencies
 are found. In this case a LO calculation seems to be more
 stable than the NLO calculation. Moreover, large NLO corrections
 are found.

 Generally, the use of $E_T^2$ as renormalisation scale is
 favoured over $Q^2$, since scale dependencies are less
 pronounced and NLO corrections are smaller.

 \vspace{-0.5cm}
 \section*{Acknowledgments}
 \vspace{-0.2cm}
 I would like to thank my colleagues P. P\"otter, T. Sch\"orner, 
 M.H. Seymour and D. Soper for the critical reading of the manuscript.

\vspace{-0.5cm}
\bibliography{jet,h1,mc,th}

%%%%%%%%
%\begin{thebibliography}{99}
%\bibitem{HERA96}
%  Proceedings `Future Physics at HERA', Eds.\ G.\ Ingelman, A.\ De Roeck, 
%  R.\ Klanner, 
%\linebreak 
%  DESY 96-235,  http://www.desy.de/\~ \protect{heraws96}
%\bibitem{www}
% http://www.desy.de/\~ \protect{heramc}/proceedings \\
%  provides instructions and templet files for contributors and conveners
%  as well as the 
%  \linebreak
%  structure of the proceedings
%\bibitem{email}
%  E-mail to workshop organizers:  heramc@desy.de
%\bibitem{label1}
%  A. Author, \Journal{\PLB}{volume}{page}{year} 
%\end{thebibliography}
\end{document}